\documentclass[12pt,reqno]{amsart}
\usepackage{graphicx}
\usepackage{amscd}
\usepackage{amsmath}
\usepackage{epsfig}
\usepackage{amsfonts}
\usepackage{amssymb}

\providecommand{\U}[1]{\protect\rule{.1in}{.1in}}
\providecommand{\U}[1]{\protect\rule{.1in}{.1in}}
\allowdisplaybreaks

\theoremstyle{plain}

\numberwithin{equation}{section}

\input{tcilatex}

\begin{document}
\title[Diffusion-Type Equation]{The Riccati Differential Equation \\
and a Diffusion-Type Equation}
\author{Erwin Suazo}
\address{School of Mathematics and Statistics, Arizona State University,
Tempe, AZ 85287--1804, U.S.A.}
\email{suazo@mathpost.la.asu.edu}
\author{Sergei K. Suslov}
\address{School of Mathematics and Statistics, Arizona State University,
Tempe, AZ 85287--1804, U.S.A.}
\email{sks@asu.edu}
\urladdr{http://hahn.la.asu.edu/\symbol{126}suslov/index.html}
\author{Jos\'{e} M. Vega-Guzm\'{a}n}
\address{School of Mathematics and Statistics, Arizona State University,
Tempe, AZ 85287--1804, U.S.A.}
\email{vega@mathpost.asu.edu}
\date{\today }

\begin{abstract}
We construct an explicit solution of the Cauchy initial value problem for
certain diffusion-type equations with variable coefficients on the entire
real line. The corresponding Green function (heat kernel) is given in terms
of elementary functions and certain integrals involving a characteristic
function, which should be found as an analytic or numerical solution of the
second order linear differential equation with time-dependent coefficients.
Some special and limiting cases are outlined. Solution of the corresponding
non-homogeneous equation is also found.
\end{abstract}

\maketitle

\section{Introduction}

In this paper we discuss explicit solution of the Cauchy initial value
problem for the one-dimensional heat equation on the entire real line%
\begin{equation}
\frac{\partial u}{\partial t}=Q\left( \frac{\partial }{\partial x},\ x,\
t\right) u,  \label{i1}
\end{equation}%
where the right hand side is a quadratic form $Q\left( p,x\right) $ of the
coordinate $x$ and the operator of differentiation $p=\partial /\partial x$
with time-dependent coefficients; see equation (\ref{h1}) below. The case of
a corresponding Schr\"{o}dinger equation is investigated in \cite%
{Cor-Sot:Lop:Sua:Sus}. In this approach, several exactly solvable models are
classified in terms of elementary solutions of a characterization equation
given by (\ref{h13}) below. Solution of the corresponding non-homogeneous
equation is obtained with the help of the Duhamel principle. These exactly
solvable cases may be of interest in a general treatment of the nonlinear
evolution equations; see \cite{Cann}, \cite{Caz}, \cite{Caz:Har}, \cite{Tao}
and references therein. Moreover, these explicit solutions can also be
useful when testing numerical methods of solving the semilinear heat
equations with variable coefficients.

\section{Solution of a Cauchy Initial Value Problem: Summary of Results}

The fundamental solution of the diffusion-type equation of the form%
\begin{equation}
\frac{\partial u}{\partial t}=a\left( t\right) \frac{\partial ^{2}u}{%
\partial x^{2}}-b\left( t\right) x^{2}u+c\left( t\right) x\frac{\partial u}{%
\partial x}+d\left( t\right) u+f\left( t\right) xu-g\left( t\right) \frac{%
\partial u}{\partial x},  \label{h1}
\end{equation}%
where $a\left( t\right) ,$ $b\left( t\right) ,$ $c\left( t\right) ,$ $%
d\left( t\right) ,$ $f\left( t\right) ,$ and $g\left( t\right) $ are given
real-valued functions of time $t$ only, can be found by a familiar
substitution%
\begin{equation}
u=Ae^{S}=A\left( t\right) e^{S\left( x,y,t\right) }  \label{h2}
\end{equation}%
with%
\begin{equation}
A=A\left( t\right) =\frac{1}{\sqrt{2\pi \mu \left( t\right) }}  \label{h3}
\end{equation}%
and%
\begin{equation}
S=S\left( x,y,t\right) =\alpha \left( t\right) x^{2}+\beta \left( t\right)
xy+\gamma \left( t\right) y^{2}+\delta \left( t\right) x+\varepsilon \left(
t\right) y+\kappa \left( t\right) ,  \label{h4}
\end{equation}%
where $\alpha \left( t\right) ,$ $\beta \left( t\right) ,$ $\gamma \left(
t\right) ,$ $\delta \left( t\right) ,$ $\varepsilon \left( t\right) ,$ and $%
\kappa \left( t\right) $ are differentiable real-valued functions of time $t$
only. Indeed,%
\begin{equation}
\frac{\partial S}{\partial t}=a\left( \frac{\partial S}{\partial x}\right)
^{2}-bx^{2}+fx+\left( cx-g\right) \frac{\partial S}{\partial x}  \label{h5}
\end{equation}%
provided%
\begin{equation}
\frac{\mu ^{\prime }}{2\mu }=-a\frac{\partial ^{2}S}{\partial x^{2}}%
-d=-2\alpha \left( t\right) a\left( t\right) -d\left( t\right) .  \label{h6}
\end{equation}%
Equating the coefficients of all admissible powers of $x^{m}y^{n}$ with $%
0\leq m+n\leq 2,$ gives the following system of ordinary differential
equations%
\begin{align}
& \frac{d\alpha }{dt}+b\left( t\right) -2c\left( t\right) \alpha -4a\left(
t\right) \alpha ^{2}=0,  \label{h7} \\
& \frac{d\beta }{dt}-\left( c\left( t\right) +4a\left( t\right) \alpha
\left( t\right) \right) \beta =0,  \label{h8} \\
& \frac{d\gamma }{dt}-a\left( t\right) \beta ^{2}\left( t\right) =0,
\label{h9} \\
& \frac{d\delta }{dt}-\left( c\left( t\right) +4a\left( t\right) \alpha
\left( t\right) \right) \delta =f\left( t\right) -2\alpha \left( t\right)
g\left( t\right) ,  \label{h10} \\
& \frac{d\varepsilon }{dt}+\left( g\left( t\right) -2a\left( t\right) \delta
\left( t\right) \right) \beta \left( t\right) =0,  \label{h11} \\
& \frac{d\kappa }{dt}+g\left( t\right) \delta \left( t\right) -a\left(
t\right) \delta ^{2}\left( t\right) =0,  \label{h12}
\end{align}%
where the first equation is the familiar Riccati nonlinear differential
equation; see, for example, \cite{Haah:Stein}, \cite{Molch}, \cite{Rainville}%
, \cite{Rajah:Mah}, \cite{Wa} and references therein.

We have%
\begin{equation*}
4a\alpha ^{\prime }+4ab-2c\left( 4a\alpha \right) -\left( 4a\alpha \right)
^{2}=0,\qquad 4a\alpha =-2d-\frac{\mu ^{\prime }}{\mu }
\end{equation*}%
from (\ref{h7}) and (\ref{h6}) and the substitution%
\begin{equation*}
4a\alpha ^{\prime }=-2d^{\prime }-\frac{\mu ^{\prime \prime }}{\mu }+\left( 
\frac{\mu ^{\prime }}{\mu }\right) ^{2}+\frac{a^{\prime }}{a}\left( 2d+\frac{%
\mu ^{\prime }}{\mu }\right)
\end{equation*}%
results in the second order linear equation%
\begin{equation}
\mu ^{\prime \prime }-\tau \left( t\right) \mu ^{\prime }-4\sigma \left(
t\right) \mu =0  \label{h13}
\end{equation}%
with%
\begin{equation}
\tau \left( t\right) =\frac{a^{\prime }}{a}+2c-4d,\qquad \sigma \left(
t\right) =ab+cd-d^{2}+\frac{d}{2}\left( \frac{a^{\prime }}{a}-\frac{%
d^{\prime }}{d}\right) .  \label{h14}
\end{equation}%
As we shall see later, equation (\ref{h13}) must be solved subject to the
initial data%
\begin{equation}
\mu \left( 0\right) =0,\qquad \mu ^{\prime }\left( 0\right) =2a\left(
0\right) \neq 0  \label{h15}
\end{equation}%
in order to satisfy the initial condition for the corresponding Green
function; see the asymptotic formula (\ref{h24}) below for a motivation.
Then, the Riccati equation (\ref{h7}) can be solved by the back substitution
(\ref{h6}).

We shall refer to equation (\ref{h13}) as the \textit{characteristic equation%
} and its solution $\mu \left( t\right) ,$ subject to (\ref{h15}), as the 
\textit{characteristic function.} As the special case (\ref{h13}) contains
the generalized equation of hypergeometric type, whose solutions are studied
in detail in \cite{Ni:Uv}; see also \cite{An:As:Ro}, \cite{Ni:Su:Uv}, \cite%
{Sus:Trey}, and \cite{Wa}.\smallskip

Thus, the Green function (fundamental solution or heat kernel) is explicitly
given in terms of the characteristic function%
\begin{equation}
u=K\left( x,y,t\right) =\frac{1}{\sqrt{2\pi \mu \left( t\right) }}\
e^{\alpha \left( t\right) x^{2}+\beta \left( t\right) xy+\gamma \left(
t\right) y^{2}+\delta \left( t\right) x+\varepsilon \left( t\right) y+\kappa
\left( t\right) }.  \label{h16}
\end{equation}%
Here%
\begin{equation}
\alpha \left( t\right) =-\frac{1}{4a\left( t\right) }\frac{\mu ^{\prime
}\left( t\right) }{\mu \left( t\right) }-\frac{d\left( t\right) }{2a\left(
t\right) },  \label{h17}
\end{equation}%
\begin{equation}
\beta \left( t\right) =\frac{1}{\mu \left( t\right) }\ \exp \left(
\int_{0}^{t}\left( c\left( \tau \right) -2d\left( \tau \right) \right) \
d\tau \right) ,  \label{h18}
\end{equation}%
\begin{eqnarray}
\gamma \left( t\right) &=&-\frac{a\left( t\right) }{\mu \left( t\right) \mu
^{\prime }\left( t\right) }\ \exp \left( 2\int_{0}^{t}\left( c\left( \tau
\right) -2d\left( \tau \right) \right) \ d\tau \right)  \label{h19} \\
&&\quad -4\int_{0}^{t}\frac{a\left( \tau \right) \sigma \left( \tau \right) 
}{\left( \mu ^{\prime }\left( \tau \right) \right) ^{2}}\left( \exp \left(
2\int_{0}^{\tau }\left( c\left( \lambda \right) -2d\left( \lambda \right)
\right) \ d\lambda \right) \right) \ d\tau ,  \notag
\end{eqnarray}%
\begin{eqnarray}
\delta \left( t\right) &=&\frac{1}{\mu \left( t\right) }\ \exp \left(
\int_{0}^{t}\left( c\left( \tau \right) -2d\left( \tau \right) \right) \
d\tau \right) \ \int_{0}^{t}\exp \left( -\int_{0}^{\tau }\left( c\left(
\lambda \right) -2d\left( \lambda \right) \right) \ d\lambda \right)  \notag
\\
&&\quad \times \left( \left( f\left( \tau \right) +\frac{d\left( \tau
\right) }{a\left( \tau \right) }g\left( \tau \right) \right) \mu \left( \tau
\right) +\frac{g\left( \tau \right) }{2a\left( \tau \right) }\mu ^{\prime
}\left( \tau \right) \right) \ d\tau ,  \label{h20}
\end{eqnarray}%
\begin{eqnarray}
\varepsilon \left( t\right) &=&-\frac{2a\left( t\right) }{\mu ^{\prime
}\left( t\right) }\delta \left( t\right) \ \exp \left( \int_{0}^{t}\left(
c\left( \tau \right) -2d\left( \tau \right) \right) \ d\tau \right)
\label{h21} \\
&&-8\int_{0}^{t}\frac{a\left( \tau \right) \sigma \left( \tau \right) }{%
\left( \mu ^{\prime }\left( \tau \right) \right) ^{2}}\exp \left(
\int_{0}^{\tau }\left( c\left( \lambda \right) -2d\left( \lambda \right)
\right) \ d\lambda \right) \left( \mu \left( \tau \right) \delta \left( \tau
\right) \right) \ d\tau  \notag \\
&&\quad +2\int_{0}^{t}\frac{a\left( \tau \right) }{\mu ^{\prime }\left( \tau
\right) }\exp \left( \int_{0}^{\tau }\left( c\left( \lambda \right)
-2d\left( \lambda \right) \right) \ d\lambda \right) \left( f\left( \tau
\right) +\frac{d\left( \tau \right) }{a\left( \tau \right) }g\left( \tau
\right) \right) \ d\tau ,  \notag
\end{eqnarray}%
\begin{eqnarray}
\kappa \left( t\right) &=&-\frac{a\left( t\right) \mu \left( t\right) }{\mu
^{\prime }\left( t\right) }\delta ^{2}\left( t\right) -4\int_{0}^{t}\frac{%
a\left( \tau \right) \sigma \left( \tau \right) }{\left( \mu ^{\prime
}\left( \tau \right) \right) ^{2}}\left( \mu \left( \tau \right) \delta
\left( \tau \right) \right) ^{2}\ d\tau  \label{h22} \\
&&\quad +2\int_{0}^{t}\frac{a\left( \tau \right) }{\mu ^{\prime }\left( \tau
\right) }\left( \mu \left( \tau \right) \delta \left( \tau \right) \right)
\left( f\left( \tau \right) +\frac{d\left( \tau \right) }{a\left( \tau
\right) }g\left( \tau \right) \right) \ d\tau  \notag
\end{eqnarray}%
with%
\begin{equation}
\delta \left( 0\right) =\frac{g\left( 0\right) }{2a\left( 0\right) },\qquad
\varepsilon \left( 0\right) =-\delta \left( 0\right) ,\qquad \kappa \left(
0\right) =0.  \label{h23}
\end{equation}%
We have used integration by parts in order to resolve the singularities of
the initial data; see section~3 for more details. Then the corresponding
asymptotic formula is%
\begin{equation}
K\left( x,y,t\right) =\frac{e^{S\left( x,y,t\right) }}{\sqrt{2\pi \mu \left(
t\right) }}\sim \frac{1}{\sqrt{4\pi a\left( 0\right) t}}\exp \left( -\frac{%
\left( x-y\right) ^{2}}{4a\left( 0\right) t}\right) \exp \left( \frac{%
g\left( 0\right) }{2a\left( 0\right) }\left( x-y\right) \right)  \label{h24}
\end{equation}%
as $t\rightarrow 0^{+}.$ Notice that the first term on the right hand side
is a familiar heat kernel for the diffusion equation with constant
coefficients (cf.~Eq.~(\ref{sp2}) below).\smallskip

By the superposition principle, we obtain solution of the Cauchy initial
value problem%
\begin{equation}
\frac{\partial u}{\partial t}=Qu,\qquad \left. u\left( x,t\right)
\right\vert _{t=0}=u_{0}\left( x\right)  \label{h25}
\end{equation}%
on the infinite interval $-\infty <x<\infty $ with the general quadratic
form $Q\left( p,x\right) $ in (\ref{h1}) as follows%
\begin{equation}
u\left( x,t\right) =\int_{-\infty }^{\infty }K\left( x,y,t\right) \
u_{0}\left( y\right) \ dy=Hu\left( x,0\right) .  \label{h26}
\end{equation}%
This yields solution explicitly in terms of an integral operator $H$ acting
on the initial data provided that the integral converges and one can
interchange differentiation and integration. This integral is essentially
the Laplace transform.

In a more general setting, solution of the initial value problem at time $%
t_{0}$ 
\begin{equation}
\frac{\partial u}{\partial t}=Qu,\qquad \left. u\left( x,t\right)
\right\vert _{t=t_{0}}=u\left( x,t_{0}\right)  \label{h27}
\end{equation}%
on an infinite interval has the form%
\begin{equation}
u\left( x,t\right) =\int_{-\infty }^{\infty }K\left( x,y,t,t_{0}\right) \
u_{0}\left( y,t_{0}\right) \ dy=H\left( t,t_{0}\right) u\left( x,t_{0}\right)
\label{h28}
\end{equation}%
with the heat kernel given by%
\begin{equation}
K\left( x,y,t,t_{0}\right) =\frac{1}{\sqrt{2\pi \mu \left( t,t_{0}\right) }}%
\ e^{\alpha \left( t,t_{0}\right) x^{2}+\beta \left( t,t_{0}\right)
xy+\gamma \left( t,t_{0}\right) y^{2}+\delta \left( t,t_{0}\right)
x+\varepsilon \left( t,t_{0}\right) y+\kappa \left( t,t_{0}\right) }.
\label{h29}
\end{equation}%
The function $\mu \left( t\right) =\mu \left( t,t_{0}\right) $ is a solution
of the characteristic equation (\ref{h13}) corresponding to the initial data%
\begin{equation}
\mu \left( t_{0},t_{0}\right) =0,\qquad \mu ^{\prime }\left(
t_{0},t_{0}\right) =2a\left( t_{0}\right) \neq 0.  \label{h30}
\end{equation}%
If $\left\{ \mu _{1},\mu _{2}\right\} $ is a fundamental solution set of
equation (\ref{h13}), then%
\begin{equation}
\mu \left( t,t_{0}\right) =\frac{2a\left( t_{0}\right) }{W\left( \mu
_{1},\mu _{2}\right) }\left( \mu _{1}\left( t_{0}\right) \mu _{2}\left(
t\right) -\mu _{1}\left( t\right) \mu _{2}\left( t_{0}\right) \right)
\label{h31}
\end{equation}%
and%
\begin{equation}
\mu ^{\prime }\left( t,t_{0}\right) =\frac{2a\left( t_{0}\right) }{W\left(
\mu _{1},\mu _{2}\right) }\left( \mu _{1}\left( t_{0}\right) \mu
_{2}^{\prime }\left( t\right) -\mu _{1}^{\prime }\left( t\right) \mu
_{2}\left( t_{0}\right) \right) ,  \label{h32}
\end{equation}%
where $W\left( \mu _{1},\mu _{2}\right) $ is the value of the Wronskian at
the point $t_{0}.$

Equations (\ref{h17})--(\ref{h22}) are valid again but with the new
characteristic function $\mu \left( t,t_{0}\right) .$ The lower limits of
integration should be replaced by $t_{0}.$ Conditions (\ref{h23}) become%
\begin{equation}
\delta \left( t_{0},t_{0}\right) =-\varepsilon \left( t_{0},t_{0}\right) =%
\frac{g\left( t_{0}\right) }{2a\left( t_{0}\right) },\qquad \kappa \left(
t_{0},t_{0}\right) =0  \label{h33}
\end{equation}%
and the asymptotic formula (\ref{h24}) should be modified as follows%
\begin{eqnarray}
K\left( x,y,t,t_{0}\right) &=&\frac{e^{S\left( x,y,t,t_{0}\right) }}{\sqrt{%
2\pi \mu \left( t,t_{0}\right) }}  \label{h34} \\
&\sim &\frac{1}{\sqrt{4\pi a\left( t_{0}\right) \left( t-t_{0}\right) }}\exp
\left( -\frac{\left( x-y\right) ^{2}}{4a\left( t_{0}\right) \left(
t-t_{0}\right) }\right) \exp \left( \frac{g\left( t_{0}\right) }{2a\left(
t_{0}\right) }\left( x-y\right) \right) .  \notag
\end{eqnarray}%
We leave the details to the reader.

\section{Derivation of The Heat Kernel}

Here we obtain the above formulas (\ref{h17})--(\ref{h22}) for the heat
kernel. The first equation is a direct consequence of (\ref{h6}) and our
equation (\ref{h8}) takes the form%
\begin{equation}
\left( \mu \beta \right) ^{\prime }=\left( c-2d\right) \left( \mu \beta
\right) ,  \label{hk1}
\end{equation}%
whose particular solution is (\ref{h18}).

From (\ref{h9}) and (\ref{h18}) one gets%
\begin{equation}
\gamma \left( t\right) =\int \frac{a\left( t\right) }{\mu ^{2}\left(
t\right) }e^{2h\left( t\right) }\ dt,\qquad h\left( t\right)
=\int_{0}^{t}\left( c\left( \tau \right) -2d\left( \tau \right) \right) \
d\tau  \label{hk2}
\end{equation}%
and integrating by parts%
\begin{equation}
\gamma \left( t\right) =-\int \frac{ae^{2h}}{\mu ^{\prime }}\ d\left( \frac{1%
}{\mu }\right) =-\frac{ae^{2h}}{\mu \mu ^{\prime }}+\int \left( \frac{ae^{2h}%
}{\mu ^{\prime }}\right) ^{\prime }\frac{dt}{\mu }.  \label{hk3}
\end{equation}%
But the derivative of the auxiliary function%
\begin{equation}
F\left( t\right) =\frac{a\left( t\right) }{\mu ^{\prime }\left( t\right) }\
e^{2h\left( t\right) }  \label{hk4}
\end{equation}%
is%
\begin{equation}
F^{\prime }\left( t\right) =\frac{\left( a^{\prime }+2h^{\prime }a\right)
e^{2h}\mu ^{\prime }-ae^{2h}\mu ^{\prime \prime }}{\left( \mu ^{\prime
}\right) ^{2}}=-\frac{4\sigma a\mu }{\left( \mu ^{\prime }\right) ^{2}}%
e^{2h}=-\frac{4\sigma \mu }{\mu ^{\prime }}F  \label{hk5}
\end{equation}%
in view of the characteristic equation (\ref{h13})--(\ref{h14}).
Substitution into (\ref{hk3}) results in (\ref{h19}).

Equation (\ref{h10}) can be rewritten as%
\begin{equation}
\left( \mu e^{-h}\delta \right) ^{\prime }=\mu e^{-h}\left( f-2\alpha
g\right) ,\qquad h=\int_{0}^{t}\left( c-2d\right) \ d\tau  \label{hk6}
\end{equation}%
and its direct integration gives (\ref{h20}).

We introduce another auxiliary function%
\begin{equation}
G\left( t\right) =\mu \left( t\right) \delta \left( t\right) e^{-h\left(
t\right) }  \label{hk7}
\end{equation}%
with the derivative given by (\ref{hk6}). Then equation (\ref{h11}) becomes 
\begin{equation*}
\frac{d\varepsilon }{dt}=-\frac{g}{\mu }e^{h}+\frac{2a\delta }{\mu }e^{h}
\end{equation*}%
and%
\begin{equation}
\varepsilon \left( t\right) =-\int \frac{g}{\mu }e^{h}\ dt+2\int \frac{aG}{%
\mu ^{2}}e^{2h}\ dt.  \label{hk8}
\end{equation}%
Integrating the second term by parts one gets%
\begin{eqnarray}
\int \frac{aG}{\mu ^{2}}e^{2h}\ dt &=&-\int \frac{aG}{\mu ^{\prime }}e^{2h}\
d\left( \frac{1}{\mu }\right) =-\int FG\ d\left( \frac{1}{\mu }\right)
\label{hk9} \\
&=&-\frac{FG}{\mu }+\int \frac{\left( FG\right) ^{\prime }}{\mu }\ dt, 
\notag
\end{eqnarray}%
where%
\begin{eqnarray}
\left( FG\right) ^{\prime } &=&F^{\prime }G+FG^{\prime }  \label{hk10} \\
&=&-\frac{4a\sigma \mu }{\left( \mu ^{\prime }\right) ^{2}}\left( \mu \delta
\right) e^{h}+\frac{a\mu }{\mu ^{\prime }}e^{h}f+\frac{d\mu }{\mu ^{\prime }}%
e^{h}g+\frac{1}{2}ge^{h}  \notag
\end{eqnarray}%
in view of (\ref{hk5}) and (\ref{hk6}). Then substitution (\ref{hk10}) into (%
\ref{hk9}) allows to cancel the divergent integrals. As a result one can
resolve the singularity and simplify expression (\ref{hk8}) to its final
form (\ref{h21}).

Finally, by (\ref{h12}) and (\ref{hk7})%
\begin{equation}
\kappa \left( t\right) =-\int g\delta \ dt+\int \frac{aG^{2}}{\mu ^{2}}%
e^{2h}\ dt,  \label{hk11}
\end{equation}%
where the last integral can be transformed as follows%
\begin{equation}
\int \frac{aG^{2}}{\mu ^{2}}e^{2h}\ dt=-\int FG^{2}\ d\left( \frac{1}{\mu }%
\right) =-\frac{FG^{2}}{\mu }+\int \frac{\left( FG^{2}\right) ^{\prime }}{%
\mu }\ dt  \label{hk12}
\end{equation}%
with%
\begin{eqnarray}
\left( FG^{2}\right) ^{\prime } &=&F^{\prime }G^{2}+2FGG^{\prime }=\left(
FG\right) ^{\prime }G+FGG^{\prime }  \label{hk13} \\
&=&-\frac{4a\sigma \mu }{\left( \mu ^{\prime }\right) ^{2}}\left( \mu \delta
\right) ^{2}+\frac{2a\mu }{\mu ^{\prime }}\left( \mu \delta \right) f+\frac{%
2d\mu }{\mu ^{\prime }}\left( \mu \delta \right) g+\mu g\delta .  \notag
\end{eqnarray}%
Substitution (\ref{hk12})--(\ref{hk13}) into (\ref{hk11}) gives our final
expression (\ref{h22}).

The details of derivation of the asymptotic formula (\ref{h24}) are left to
the reader.

\section{Special Initial Data}

In the case $u\left( x,0\right) =u_{0}=$constant$,$ our solution (\ref{h26})
takes the form%
\begin{eqnarray}
u\left( x,t\right) &=&\int_{-\infty }^{\infty }K\left( x,y,t\right) \ u_{0}\
dy  \label{sid1} \\
&=&u_{0}\frac{e^{\alpha \left( t\right) x^{2}+\delta \left( t\right)
x+\kappa \left( t\right) }}{\sqrt{2\pi \mu \left( t\right) }}\ \int_{-\infty
}^{\infty }e^{\left( \beta \left( t\right) x+\varepsilon \left( t\right)
\right) y+\gamma \left( t\right) y^{2}}\ dy  \notag \\
&=&\frac{u_{0}}{\sqrt{-2\mu \gamma }}\exp \left( \frac{\left( 4\alpha \gamma
-\beta ^{2}\right) x^{2}+2\left( 2\gamma \delta -\beta \varepsilon \right)
x+4\gamma \kappa -\varepsilon ^{2}}{4\gamma }\right) ,  \notag
\end{eqnarray}%
provided $\gamma \left( t\right) <0,$ with the help of an elementary integral%
\begin{equation}
\int_{-\infty }^{\infty }e^{-ay^{2}+2by}\ dy=\sqrt{\frac{\pi }{a}}\
e^{b^{2}/a},\qquad a>0.  \label{sid2}
\end{equation}%
The details of taking the limit $t\rightarrow 0^{+}$ in (\ref{sid1}) are
left to the reader.

When $u\left( x,0\right) =\delta \left( x-x_{0}\right) ,$ where $\delta
\left( x\right) $ is the Dirac delta function, one gets formally%
\begin{equation}
u\left( x,t\right) =\int_{-\infty }^{\infty }K\left( x,y,t\right) \ \delta
\left( y-x_{0}\right) \ dy=K\left( x,x_{0},t\right) .  \label{sid3}
\end{equation}%
Thus, in general, the heat kernel (\ref{h16}) provides an evolution of this
initial data, concentrated originally at a point $x_{0},$ into the entire
space for a suitable time interval $t>0.$

\section{Some Examples}

Now let us consider several elementary solutions of the characteristic
equation (\ref{h13}); more complicated cases may include special functions,
like Bessel, hypergeometric or elliptic functions \cite{An:As:Ro}, \cite%
{Ni:Uv}, \cite{Rain}, and \cite{Wa}. Among important elementary cases of our
general expressions for the Green function (\ref{h16})--(\ref{h22}) are the
following:\smallskip

For the traditional diffusion equation%
\begin{equation}
\frac{\partial u}{\partial t}=a\frac{\partial ^{2}u}{\partial x^{2}},\qquad
a=\text{constant}>0  \label{sp1}
\end{equation}%
the heat kernel is%
\begin{equation}
K\left( x,y,t\right) =\frac{1}{\sqrt{4\pi at}}\exp \left( -\frac{\left(
x-y\right) ^{2}}{4at}\right) ,\qquad t>0.  \label{sp2}
\end{equation}%
Equation (\ref{sid1}) gives the steady solution $u_{0}=$constant for all
times $t\geq 0.$ See \cite{Cann} and references therein for a detailed
investigation of the classical one-dimensional heat equation.

The diffusion-type equation%
\begin{equation}
\frac{\partial u}{\partial t}=a\frac{\partial ^{2}u}{\partial x^{2}}+fxu,
\label{sp1a}
\end{equation}%
where $a>0$ and $f$ are constants (see \cite{FeynmanPhD}, \cite{Feynman}, 
\cite{Feynman49a}, \cite{Feynman49b}, \cite{Fey:Hib}, \cite%
{Cor-Sot:Lop:Sua:Sus} and references therein regarding to similar cases of
the Schr\"{o}dinger equation), has the the characteristic function of the
form $\mu =2at.$ The heat kernel is%
\begin{equation}
K\left( x,y,t\right) =\frac{1}{\sqrt{4\pi at}}\exp \left( -\frac{\left(
x-y\right) ^{2}}{4at}\right) \exp \left( \frac{f}{2}\left( x+y\right) t+%
\frac{af^{2}}{12}t^{3}\right)  \label{sp2a}
\end{equation}%
provided $t>0.$ Evolution of the uniform initial data $u\left( x,0\right)
=u_{0}=$constant is given by%
\begin{equation}
u\left( x,t\right) =u_{0}e^{fxt+af^{2}t^{3}/3}.  \label{sp2b}
\end{equation}

The initial value problem for the following diffusion-type equation with
variable coefficients%
\begin{equation}
\frac{\partial u}{\partial t}=a\left( \frac{\partial ^{2}u}{\partial x^{2}}%
-x^{2}u\right) +\omega \left( \cosh \left( \left( 2a-1\right) t\right) \
xu+\sinh \left( \left( 2a-1\right) t\right) \ \frac{\partial u}{\partial x}%
\right) ,  \label{diff1}
\end{equation}%
where $a>0$ and $\omega $ are two constants, was solved in \cite{Lop:Sus} by
using the eigenfunction expansion method and a connection with the
representations of the Heisenberg--Weyl group $N\left( 3\right) .$ Here we
apply a different approach. The solution of the characteristic equation%
\begin{equation}
\mu ^{\prime \prime }-4a^{2}\mu =0  \label{diff2}
\end{equation}%
is $\mu =\sinh \left( 2at\right) $ and the corresponding heat kernel is
given by%
\begin{align}
& K\left( x,y,t\right) =\frac{1}{\sqrt{2\pi \sinh \left( 2at\right) }}\ \exp
\left( -\frac{\left( x^{2}+y^{2}\right) \cosh \left( 2at\right) -2xy}{2\sinh
\left( 2at\right) }\right)  \label{diff3a} \\
& \qquad \qquad \quad \times \exp \left( 2\omega \frac{x\sinh \left(
t/2\right) +y\sinh \left( \left( 2a-1/2\right) t\right) }{\sinh \left(
2at\right) }\sinh \left( \frac{t}{2}\right) \right)  \notag \\
& \qquad \qquad \qquad \times \exp \left( -2\omega ^{2}\frac{\cosh \left(
2at\right) }{\sinh \left( 2at\right) }\sinh ^{4}\left( \frac{t}{2}\right)
\right)  \notag  \label{diff3} \\
& \qquad \qquad \quad \quad \quad \times \exp \left( \frac{\omega ^{2}}{2}%
\left( t-2\sinh t+\frac{1}{2}\sinh \left( 2t\right) \right) \right) ,\qquad
t>0.  \notag
\end{align}%
Indeed, by (\ref{h17})--(\ref{h19})%
\begin{equation}
\alpha =\gamma =-\frac{\cosh \left( 2at\right) }{2\sinh \left( 2at\right) }%
,\qquad \beta =\frac{1}{\sinh \left( 2at\right) }.  \label{diff3b}
\end{equation}%
In this case%
\begin{eqnarray*}
&&f\mu +\frac{g}{2a}\mu ^{\prime } \\
&&\ =\omega \left( \cosh \left( \left( 2a-1\right) t\right) \sinh \left(
2at\right) -\sinh \left( \left( 2a-1\right) t\right) \cosh \left( 2at\right)
\right) \\
&&\ =\omega \sinh t
\end{eqnarray*}%
and equation (\ref{h20}) gives%
\begin{equation}
\delta =\omega \frac{\cosh t-1}{\sinh \left( 2at\right) }=2\omega \frac{%
\sinh ^{2}\left( t/2\right) }{\sinh \left( 2at\right) }.  \label{diff3c}
\end{equation}%
By (\ref{h21})%
\begin{eqnarray}
\varepsilon &=&\omega \frac{1-\cosh t}{\sinh \left( 2at\right) \cosh \left(
2at\right) }  \label{diff4} \\
&&+2a\omega \int_{0}^{t}\frac{1-\cosh \tau }{\cosh ^{2}\left( 2a\tau \right) 
}\ d\tau +\omega \int_{0}^{t}\frac{\cosh \left( \left( 2a-1\right) \tau
\right) }{\cosh \left( 2a\tau \right) }\ d\tau ,  \notag
\end{eqnarray}%
where the integration by parts gives%
\begin{equation*}
2a\int_{0}^{t}\frac{1-\cosh \tau }{\cosh ^{2}\left( 2a\tau \right) }\ d\tau
=\left( 1-\cosh t\right) \frac{\sinh \left( 2at\right) }{\cosh \left(
2at\right) }+\int_{0}^{t}\frac{\sinh \left( 2a\tau \right) }{\cosh \left(
2a\tau \right) }\sinh \tau \ d\tau .
\end{equation*}%
Thus%
\begin{equation*}
\varepsilon =\omega \left( 1-\cosh t\right) \frac{\cosh \left( 2at\right) }{%
\sinh \left( 2at\right) }+\omega \int_{0}^{t}\frac{\sinh \left( 2a\tau
\right) \sinh \tau +\cosh \left( \left( 2a-1\right) \tau \right) }{\cosh
\left( 2a\tau \right) }\ d\tau
\end{equation*}%
and an elementary identity%
\begin{equation}
\sinh \left( 2at\right) \sinh t+\cosh \left( \left( 2a-1\right) t\right)
=\cosh \left( 2at\right) \cosh t  \label{id1}
\end{equation}%
leads to an integral evaluation. Two other identities%
\begin{eqnarray}
&&\cosh \left( 2at\right) \cosh t-\sinh \left( 2at\right) \sinh t=\cosh
\left( \left( 2a-1\right) t\right) ,  \label{id2} \\
&&\cosh \left( 2at\right) -\cosh \left( \left( 2a-1\right) t\right) =2\sinh
\left( t/2\right) \sinh \left( \left( 2a-1/2\right) t\right)  \label{id3}
\end{eqnarray}%
result in%
\begin{equation}
\varepsilon =\omega \frac{\cosh \left( 2at\right) -\cosh \left( \left(
2a-1\right) t\right) }{\sinh \left( 2at\right) }=2\omega \frac{\sinh \left(
t/2\right) \sinh \left( \left( 2a-1/2\right) t\right) }{\sinh \left(
2at\right) }.  \label{diff5}
\end{equation}%
In a similar fashion,%
\begin{equation}
\kappa =-2\omega ^{2}\sinh ^{4}\left( t/2\right) \frac{\cosh \left(
2at\right) }{\sinh \left( 2at\right) }+\frac{1}{2}\omega ^{2}\left( t-2\sinh
t+\frac{1}{2}\sinh \left( 2t\right) \right) ,  \label{diff6}
\end{equation}%
and equation (\ref{diff3a}) is derived. In the limit $\omega \rightarrow 0$
this kernel gives also a familiar expression in statistical mechanics for
the density matrix for a system consisting of a simple harmonic oscillator 
\cite{Fey:Hib}.

The case $a=1/2$ corresponds to the equation%
\begin{equation}
\frac{\partial u}{\partial t}=\frac{1}{2}\left( \frac{\partial ^{2}u}{%
\partial x^{2}}-x^{2}u\right) +\omega \ xu  \label{diff7}
\end{equation}%
and the heat kernel (\ref{diff3a}) is simplified to the form%
\begin{equation}
K\left( x,y,t\right) =\frac{e^{\omega ^{2}t/2}}{\sqrt{2\pi \sinh t}}\ \exp
\left( -\frac{\left( \left( x-\omega \right) ^{2}+\left( y-\omega \right)
^{2}\right) \cosh t-2\left( x-\omega \right) \left( y-\omega \right) }{%
2\sinh t}\right) ,  \label{diff7ab}
\end{equation}%
when $t>0.$ A similar diffusion-type equation%
\begin{equation}
\frac{\partial u}{\partial t}=\frac{1}{2}\left( \frac{\partial ^{2}u}{%
\partial x^{2}}+x^{2}u\right) +\omega \ xu  \label{diff7a}
\end{equation}%
can be solved with the aid of the kernel%
\begin{equation}
K\left( x,y,t\right) =\frac{e^{-\omega ^{2}t/2}}{\sqrt{2\pi \sin t}}\ \exp
\left( -\frac{\left( \left( x+\omega \right) ^{2}+\left( y+\omega \right)
^{2}\right) \cos t-2\left( x+\omega \right) \left( y+\omega \right) }{2\sin t%
}\right)  \label{diff8a}
\end{equation}%
provided $0<t<\pi /2.$ We leave the details to the reader.

Following to the case of exactly solvable time-dependent Schr\"{o}dinger
equation found in \cite{Me:Co:Su}, we consider the diffusion-type equation
of the form%
\begin{equation}
\frac{\partial u}{\partial t}=\cosh ^{2}t\ \frac{\partial ^{2}u}{\partial
x^{2}}+\sinh ^{2}t\ x^{2}u+\frac{1}{2}\sinh 2t\left( 2x\frac{\partial u}{%
\partial x}+u\right) .  \label{sp3}
\end{equation}%
The corresponding characteristic equation%
\begin{equation}
\mu ^{\prime \prime }-2\tanh t\ \mu ^{\prime }+2\mu =0  \label{sp4}
\end{equation}%
has two linearly independent solutions%
\begin{eqnarray}
\mu _{1} &=&\cos t\sinh t+\sin t\cosh t,  \label{sp5} \\
\mu _{2} &=&\sin t\sinh t-\cos t\cosh t  \label{sp6}
\end{eqnarray}%
with the Wronskian $W\left( \mu _{1},\mu _{2}\right) =2\cosh ^{2}t,$ and the
first one satisfies the initial conditions (\ref{h15}). The heat kernel is%
\begin{eqnarray}
K\left( x,y,t\right) &=&\frac{1}{\sqrt{2\pi \left( \cos t\sinh t+\sin t\cosh
t\right) }}  \label{sp7} \\
&&\times \exp \left( \frac{\left( y^{2}-x^{2}\right) \sin t\sinh
t+2xy-\left( x^{2}+y^{2}\right) \cos t\cosh t}{2\left( \cos t\sinh t+\sin
t\cosh t\right) }\right)  \notag
\end{eqnarray}%
provided $0<t<T_{1}\approx 0.9375520344,$ where $T_{1}$ is the first
positive root of the transcendental equation $\tanh t=\cot t.$ Then $\gamma
\left( t\right) <0$ and the integral (\ref{h26}) converges for suitable
initial data.$\allowbreak $

A similar diffusion-type equation%
\begin{equation}
\frac{\partial u}{\partial t}=\cos ^{2}t\ \frac{\partial ^{2}u}{\partial
x^{2}}+\sin ^{2}t\ x^{2}u-\frac{1}{2}\sin 2t\left( 2x\frac{\partial u}{%
\partial x}+u\right)   \label{sp8}
\end{equation}%
has the characteristic equation of the form%
\begin{equation}
\mu ^{\prime \prime }+2\tan t\ \mu ^{\prime }-2\mu =0  \label{sp9}
\end{equation}%
with the same solution (\ref{sp5}). It appeared in \cite{Me:Co:Su} and \cite%
{Cor-Sot:Lop:Sua:Sus} for a special case of the Schr\"{o}dinger equation.
The corresponding heat kernel has the same form (\ref{sp7}) but with $x$ and 
$y$ interchanged:%
\begin{eqnarray}
K\left( x,y,t\right)  &=&\frac{1}{\sqrt{2\pi \left( \cos t\sinh t+\sin
t\cosh t\right) }}  \label{sp10} \\
&&\times \exp \left( \frac{\left( x^{2}-y^{2}\right) \sin t\sinh
t+2xy-\left( x^{2}+y^{2}\right) \cos t\cosh t}{2\left( \cos t\sinh t+\sin
t\cosh t\right) }\right)   \notag
\end{eqnarray}%
provided $0<t<T_{2}\approx 2.347045566,$ where $T_{2}$ is the first positive
root of the transcendental equation $\tanh t=-\cot t.$ We leave the details
for the reader.

\section{Solution of the Non-Homogeneous Equation}

A diffusion-type equation of the form%
\begin{equation}
\left( \frac{\partial }{\partial t}-Q\left( t\right) \right) u=F,
\label{nh1}
\end{equation}%
where $Q$ stands for the second order linear differential operator in the
right hand side of equation (\ref{h1}) and $F=F\left( t,x,u\right) ,$ can be
rewritten formally as an integral equation (the Duhamel principle; see \cite%
{Caz}, \cite{Caz:Har}, \cite{Lad:Sol:Ural}, \cite{Levi}, \cite{Sua:Sus}, 
\cite{Tao} and references therein)%
\begin{equation}
u\left( x,t\right) =H\left( t,0\right) u\left( x,0\right)
+\int_{0}^{t}H\left( t,s\right) F\left( s,x,u\right) \ ds.  \label{nh2}
\end{equation}%
Operator $H\left( t,s\right) $ is given by (\ref{h28}). When $F$ does not
depend on $u,$ one gets a solution of the nonhomogeneous equation (\ref{nh1}%
).

Indeed, a formal differentiation gives%
\begin{equation}
\frac{\partial u}{\partial t}=\frac{\partial }{\partial t}H\left( t,0\right)
u\left( x,0\right) +\frac{\partial }{\partial t}\int_{0}^{t}H\left(
t,s\right) F\left( s,x,u\right) \ ds,  \label{nh3}
\end{equation}%
where%
\begin{equation}
\frac{\partial }{\partial t}\int_{0}^{t}H\left( t,s\right) \ F\left(
s,x,u\right) \ ds=H\left( t,t\right) \ F\left( t,x,u\right) +\int_{0}^{t}%
\frac{\partial }{\partial t}H\left( t,s\right) F\left( s,x,u\right) \ ds
\label{nh4}
\end{equation}%
and we assume that $H\left( t,t\right) $ is the identity operator. Also%
\begin{equation}
Q\left( t\right) u=Q\left( t\right) H\left( t,0\right) u\left( x,0\right)
+\int_{0}^{t}Q\left( t\right) H\left( t,s\right) F\left( s,x,u\right) \ ds
\label{nh5}
\end{equation}%
and%
\begin{eqnarray}
\left( \frac{\partial }{\partial t}-Q\left( t\right) \right) u &=&\left( 
\frac{\partial }{\partial t}-Q\left( t\right) \right) H\left( t,0\right)
u\left( x,0\right) +F  \label{nh6} \\
&&+\int_{0}^{t}\left( \frac{\partial }{\partial t}-Q\left( t\right) \right)
H\left( t,s\right) F\left( s,x,u\right) \ ds,  \notag
\end{eqnarray}%
where%
\begin{equation}
\left( \frac{\partial }{\partial t}-Q\left( t\right) \right) H\left(
t,s\right) =0,\qquad 0\leq s<t  \label{nh7}
\end{equation}%
by construction of the operator $H\left( t,s\right) $ in (\ref{h28}). This
completes our formal proof. A rigorous proof will be given elsewhere.

\noindent \textbf{Acknowledgments.\/} This paper is written as a part of the
summer 2008 program on analysis of Mathematical and Theoretical Biology
Institute (MTBI) at Arizona State University. The MTBI/SUMS Summer
Undergraduate Research Program is supported by The National Science
Foundation (DMS-0502349), The National Security Agency (DOD-H982300710096),
The Sloan Foundation, and Arizona State University. The authors are grateful
to Professor Carlos Castillo-Ch\'{a}vez for support and reference \cite%
{Bet:Cin:Kai:Cas}. We thank Professors Faina Berezovskaya, Alex Mahalov, and
Svetlana Roudenko for valuable comments.

\end{document}